\documentclass[10pt,pra,letterpaper,typeset,twocolumn,showpacs,floatfix]{revtex4-1}

\usepackage{color}
\usepackage{graphicx}
\usepackage{amsmath}
\usepackage{amssymb,amsthm}
\graphicspath{{pict/}{}}

\usepackage{bm}

\newcounter{Fig}

\newcommand{\be}{\begin{equation}}
\newcommand{\ee}{\end{equation}}

\begin{document}

\title{A simple method to construct Flat Band lattices}

\author{Luis Morales-Inostroza}
\affiliation{Departamento de F\'isica and MSI-Nucleus on Advanced Optics, Center for Optics and Photonics (CEFOP), Facultad de Ciencias, Universidad de Chile, Santiago, Chile}
\author{Rodrigo A. Vicencio}
\affiliation{Departamento de F\'isica and MSI-Nucleus on Advanced Optics, Center for Optics and Photonics (CEFOP), Facultad de Ciencias, Universidad de Chile, Santiago, Chile}

\pacs{42.82.Et, 63.20.Pw, 78.67.Pt, 42.65.Tg}

\begin{abstract}

We develop a simple and general method to construct arbitrary Flat Band lattices. We identify the basic ingredients behind zero-dispersion bands and develop a method to construct extended lattices based on a consecutive repetition of a given mini-array. The number of degenerated localized states is defined by the number of connected mini-arrays times the number of modes preserving the symmetry at a given connector site. In this way, we create one or more (depending on the lattice geometry) complete degenerated Flat Bands for quasi-one and two-dimensional systems. We probe our method by studying several examples, and discuss the effect of additional interactions like anisotropy or nonlinearity. At the end, we test our method by studying numerically a ribbon lattice using a continuous description.

\end{abstract}

\maketitle

\section{Introduction}

Lattices or extended periodical systems constitute a central framework for several areas of research. For example, in solid-state physics periodical lattices are a fundamental starting point~\cite{kittel}, where many general predictions have been made without a direct experimental observation. The last decades, photonic lattices have emerged as key experimental setups where to study most of the predicted electronic properties, due to a direct observation of the optical wave-function using a simple CCD camera~\cite{rep1}. In this context, Anderson Localization~\cite{anderson} was directly observed, for the first time, in two-dimensional (2D) waveguide arrays~\cite{segev} and, then, also in one-dimensional (1D) lattices~\cite{morandotti1}. In this case, the energy is trapped due to consecutive destructive interference from randomly distributed scatters (sites or waveguides). Before this very fundamental observation, localization resulting as a balance between discreteness (diffraction) and nonlinearity (self-focusing) was observed in 1D~\cite{morandotti2,segev2,alex1} and 2D~\cite{segev3,alex2Ds} lattices. Nonlinear localized modes of this nature are known as Discrete Solitons or Intrinsic Localized Modes~\cite{PT04}, and the conditions for existence and stability, in diverse contexts, are nowadays well understood~\cite{rep2}. The nonlinearity and larger intensities create effective defects regions, where the trapping potential changes locally. In this way, waves naturally get trapped at deeper potential wells. In the nonlinear case, this occurs in a perfectly periodic lattice; however, this type of phenomena can also be observed in a linear regime by directly inserting a linear defect into a homogeneous system~\cite{Chen1,Chen1a}. Exponentially decaying localized modes are obtained in both --linear and nonlinear-- cases, depending on the effective strength of the induced defect. Therefore, these modes are not compact and high localization demands a very strong effective disturbance.

Another interesting way to induce localization on a lattice originates from the understanding of the linear properties of a given unconventional array. In standard/conventional systems (e.g., square, hexagonal or honeycomb lattices) the linear spectrum is always dispersive (excepting some $k$-space points where the derivative becomes locally zero). As a result, when exciting the lattice, for example using a single-site excitation, a set of linear extended modes will be excited and waves will propagate incoherently across the lattice, depending on their specific velocities. Therefore, stationary localized patterns will be not observed for perfectly periodic linear standard lattices and the energy will diffuse only~\cite{rep1}. However, by carefully selecting the geometry of the lattice, a different kind of linear spectrum can be obtained. A Flat Band (FB) --unconventional-- lattice possesses a unique linear spectrum. In these quasi-1D or 2D systems, a complete band (not only a section of it) is completely flat, implying zero dispersion and not diffraction at all for the states belonging to this band. Diamond~\cite{ThomsonD}, Stub~\cite{Amo}, Sawtooth~\cite{chi1,SawOL}, Kagome~\cite{stan1,chen2} or Lieb~\cite{prlus,ThomsonL,sci1} lattices are some examples of recent explored FB systems, in diverse physical contexts. These examples show the diversity of fabrication techniques and impressive possibilities for creating, in principle, any wished lattice. In all these geometries, a FB is composed by a large set of degenerated localized linear modes, all of them propagating coherently along the lattice. Moreover, these states occupy only few lattice sites being exactly compact (zero tail), in a perfectly periodic linear lattice. This implies that the localization is always perfect and it does not depend on any external parameter. The special geometry of FB systems generates consecutive phase cancellations that, effectively, reduces the excited region to a mini-array of a given lattice. The linear combination of these localized states is thought to be important for applications in all-optical imaging transmission~\cite{kagome1,Chen3}, as a secure and compact mechanism of transporting information at a very low level of power. 

There are several attempts to find a simple method to construct FB lattices~\cite{origin,japan,bergman,flach}. However, we have not found a direct, simple and general method to create any wished FB system with given specific features (e.g., having several zero dispersion bands). For example, Ref.~\cite{japan} starts from a given 2D or 3D lattice, and constructs partial line graphs without much connection with the physics of the sub-lattices, which is of major relevance in our method. In Ref.~\cite{manni} authors describe a collection of FB lattices to study bosons and fermions dynamics. They briefly mention that an important condition is that the wave-function may be zero at some connecting sites, in order to make impossible the transport across the lattice. It is important to notice that this is an important requirement but it is not complete. It is possible to excite any lattice with a profile having zero amplitudes at some connecting sites; however, the energy could nevertheless diffuse and the FB will be not exclusively excited. To observe localization, as a result of the excitation of any FB, a mandatory condition is to excite the lattice by one of the modes of a given mini-array (or a linear combination of them), that has zero amplitudes at the connecting sites. The additive destructive interference will make possible the cancellation of the amplitude at the connecting sites and the transport will be just forbidden. 

In this work, we identify the key ingredients to construct arbitrary lattices possessing one or more flat bands, focusing on the consecutive addition of mini-arrays to form an extended lattice. Our method is based on the knowledge of the fundamental linear modes belonging to a given mini-array, in order to assure the cancellation of phases at connector sites. The existence of a linear mode satisfying this condition constitutes a proof to guarantee the existence of a Flat Band, which is composed of compact localized states. Different systems in quasi-one and two dimensions are generated with one or more Flat Bands (it is straightforward to extend our method to 3D as well). Additionally, we also show that it is possible to use our method to construct aperiodical systems having a set of compact modes forming a full FB. Extra considerations as, for example, anisotropy, next-nearest neighbors interactions or local nonlinearity are also discussed. At the end, we study a ribbon lattice using a continuous model where we test our discrete predictions in a more realistic configuration.

\section{General Model}

Focusing on written and induced waveguide lattices~\cite{rep1}, we model the propagation of light in weakly coupled systems using a Discrete Linear Schr\"odinger (DLS) Equation. In this model, a given waveguide mode weakly interacts with its close neighbors due to an evanescent interaction. The amplitude of the mode at the ${\vec n}$-th waveguide position is given by $\psi_{\vec n}$, and its dynamical evolution is simply modelled as
\begin{equation}\label{DLS}
-i \frac {d \psi_{\vec n}}{d z} = \epsilon_{\vec n} \psi_{\vec n} +\sum_{\vec m \not= \vec n}V_{nm} \psi_{\vec m}\ ,
\end{equation}
where $z$ is the propagation coordinate (in other contexts, it corresponds to time~\cite{rep1,rep2}). $\epsilon_{\vec n}$ corresponds to the propagation constant at the ${\vec n}$-th site (if the lattice is homogenous, we simply set $\epsilon_{\vec n}=0$, without loss of generality). $V_{nm}$ describes the coupling interaction between the ${\vec n}$-th and ${\vec m}$-th sites. When constructing the lattice, these coefficients define all the linear interactions between close sites, according to a given lattice geometry. In fact, as we will discuss below, some FBs also survive when including next nearest-neighbors interactions; of course, taking into account the exponentially decaying tendency of coupling constants with respect to separation distance~\cite{alex2}. For some FB lattices the anisotropy or next-nearest neighbors interactions destroy the flatness of the band. However, we will show that our method allows the construction of more robust systems, what is important to observe the FB phenomenology in real experiments~\cite{ThomsonD,SawOL,prlus,ThomsonL,chen2,Chen3,anton2}.

In general, the linear properties of any periodical system are contained in the definition of $V_{nm}$ coefficients. We solve the stationary problem by using a plane wave (Bloch) ansatz $$\psi_{\vec n} (z)=A_{\vec n}e^{i \vec{k} \cdot \vec{n}  }e^{i \beta z}\ .$$ By inserting this into model (\ref{DLS}), we obtain the following set of coupled stationary equations
\begin{equation}
\beta(\vec{k}) A_{\vec n} =  \sum_{m \neq n}V_{nm} A_{\vec m}\ e^{i \vec{k} \cdot (\vec{m}-\vec{n})  }\ .
\label{sta}
\end{equation}
The number of sites per unitary cell defines the number of different amplitude $A_{\vec n}$ to be considered. For example, in a system with three sites per unit cell (e.g., Stub, Lieb or Kagome lattices), we will require three different amplitudes to characterize the linear properties of the system having three linear bands. By solving the eigenvalue problem (\ref{sta}), we obtain the linear spectrum of a given lattice according to the interactions included in $V_{nm}$. Along this work, we will consider a constant spatial period of $a=1$, in order to simplify the expressions.


\section{Two-sites case}

\begin{figure}[t!] 
 \centering
   \includegraphics[width=.45\textwidth]{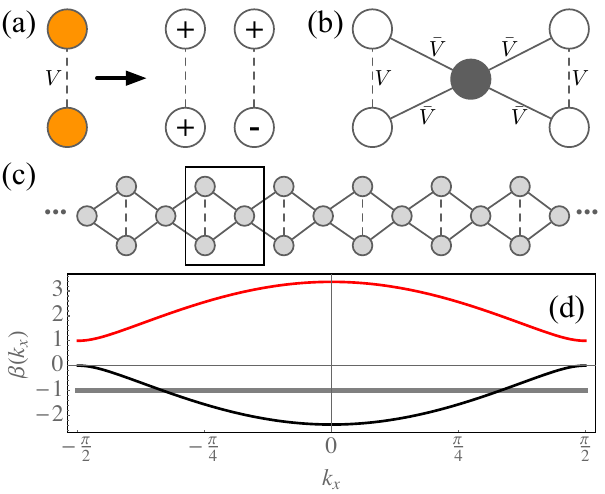}
\caption{(a) A dimer and its linear modes. (b) Two mini-arrays connected by a connector (dark) site. (c) Rhombic lattice. (d) Linear spectrum for $V=\bar{V}=1$.}\label{fig1}
\end{figure}

Let us start with the simplest possible mini-array: a dimer [see Fig.~\ref{fig1}(a)-left]. This basic system describes two close waveguides interacting via a coupling coefficient $V$. A dimer possesses two linear stationary modes: $A_1=A_2$ and $A_1=-A_2$ [symbolic notation $\{+,+\}$ and $\{+,-\}$, respectively], as shown in Fig.~\ref{fig1}(a)-right (for simplicity, we do not include any mode normalization along the text). The frequency of these modes is $\beta=V$ and $\beta=-V$, respectively. Now, we increase the system size by adding a ``connector'' site in between two vertically oriented dimers [see Fig.~\ref{fig1}(b)]. We include an additional coupling $\bar{V}$ (full line), which is determined by the specific geometry (angle and distance~\cite{alex2}). Immediately, we realize that the original dimer solution $\{+,-\}$, located in any of the two dimers, also corresponds to a solution (mode) of this new composed system. For this mode, Eqs.~(\ref{sta}) read 
%
\begin{eqnarray*}
\beta\cdot 0=\bar{V}\cdot(A-A)=0\ ,\\
\beta\cdot \pm A = V\cdot \mp A+\bar{V}\cdot0\Rightarrow \beta=-V\ .
\end{eqnarray*}
%
%
The phase difference between the non-zero amplitude sites allows the cancellation of the amplitude at the added connector site. The frequency of this mode, in the new composed system, remains equal to the one of the original dimer problem ($\beta=-V$). By adding more connector sites and dimers (mini-arrays), we are able to construct a full extended lattice, as the one shown in Fig.~\ref{fig1}(c). This lattice is known as Rhombic or Diamond lattice and has been recently studied experimentally in Ref.~\cite{ThomsonD}. As the $\{+,-\}$ mode is still a mode in the extended system, there will be as many of these modes as the number of mini-arrays (dimers) in the lattice, all of them having the same frequency $\beta=-V$. Therefore, all these modes will form a completely degenerated Flat Band. 

We solve the linear problem (\ref{sta}) for the full Rhombic lattice, considering the unitary cell formed by three sites [enclosed area in Fig.~\ref{fig1}(c)], and find the bands $$\beta(k_x)=-V,\ \left(V \pm \sqrt{V^2 + 32 \bar{V}^2 \cos^2 (k_x)}\right)/2\ .$$ We realize that the mini-array mode is also an exact solution of the extended system forming a full Flat Band at $\beta=-V$, independent on the $\bar{V}$ value. Fig.~\ref{fig1}(d) shows the linear spectrum for this Diamond lattice.

The discreteness of the system and the symmetry at the connector site equation, when considering a linear mode of the mini-array, are the keys of success to create any lattice possessing, at least, one Flat Band. For the previous example, and assuming a $\{+,-\}$ mode (or combinations of it) as an initial condition at $z=0$, the connector equation becomes
$$-i \frac {d \psi_{\vec C}}{d z} =  \sum_{m \neq C}V_{C,m} \psi_{\vec m}(z=0)=0\Rightarrow \psi_{\vec C}(z)=0\ ,$$
because $\psi_{\vec C}(z=0)=0$. Considering this example and extending it to other configurations, we could claim that we can construct any FB lattice by connecting any given mini-array (dimer, trimer, rhombus, etc.) via different connector sites in different directions, assuring that the dynamical equation for this site may be strictly equal to zero. This is achieved by using a specific mini-array mode as an initial condition, what effectively allows the cancellation of phases at the connector site. In this way, we can construct a full lattice with a zero dispersion band, where this particular mini-array mode will be a degenerated solution of the extended system. One way to assure this is, for example, by choosing the connector sites coinciding with a node of a given mode of the mini-array. Therefore, this mode will naturally become a mode of the Flat Band because it will preserve the symmetry at the dynamical connector site equation. Of course, this method is not unique, but when it is successfully applied it assures the existence of a FB.

\section{Three-sites case}

%
\begin{figure}[h!] 
 \centering
   \includegraphics[width=.4\textwidth]{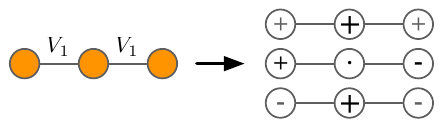}
\caption{Left: A three sites mini-array. Right: Linear modes.}\label{fig2}
\end{figure}
%
Another interesting example is based on a mini-array consisting of only three sites connected in a row, as sketched in Fig.~\ref{fig2}-left. This simple system possesses three linear stationary modes: $$\{+,\sqrt{2},+\},\ \{+,0,- \},\ \{-,\sqrt{2},-\}\ ,$$ as shown in Fig.~\ref{fig2}-right, with frequencies: $\beta=\sqrt{2},\ 0,-\sqrt{2}$, respectively. We will show that we can use two of these modes to construct two different FB lattices.

\subsection{``Cross'' lattice}

\begin{figure}[h!] 
 \centering
   \includegraphics[width=.45\textwidth]{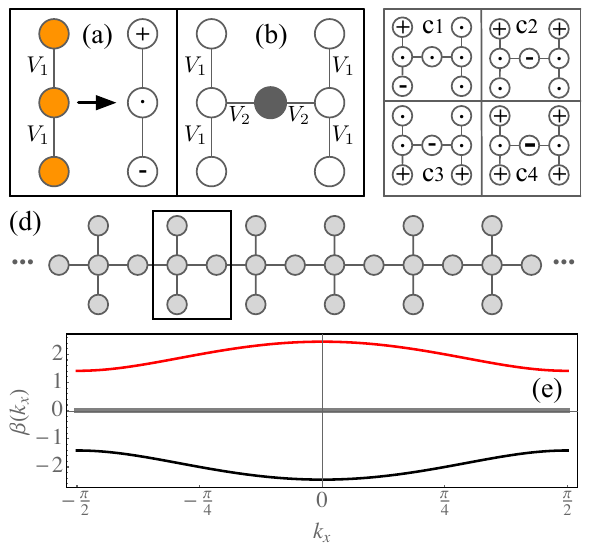}
\caption{(a) A three sites mini-array and one of its modes. (b) Two mini-arrays connected by a connector (dark) site. (c) Four linear modes of the composed system. (d) Cross lattice. (e) Linear spectrum for $V_1=V_2=1$.}\label{fgcross}
\end{figure}

We start our lattice composition by selecting one of the three linear modes of the mini-array. The first mode does not present any phase oscillation. Therefore, it will not induce cancellation of transport on a given connector site, neither it will be associated to any Flat Band (however, this could change by assuming very rear negative coupling constants~\cite{unisign}). The second $\{+,0,- \}$ mode is similar to the mode used for the dimer to create the rhombic lattice, but it has an extra zero amplitude site at the center [see Fig.~\ref{fgcross}(a)]. So, it becomes natural to use this null site to connect two mini-arrays. However, if we connect a second mini-array, for example, to the right, we would end up with a system that will not preserve the symmetry of the mini-array mode, and our method will simply not work (there is a way to do it in a higher dimension, by rotating the mini-arrays in $90^\circ$ consecutively, obtaining a 3D FB spine-like ribbon). Therefore, we add an extra connector site in between two mini-arrays as shown in Fig.~\ref{fgcross}(b), using a horizontal coupling $V_2$. As we expect, the trimer $\{+,0,- \}$ mode is also a solution of this new system [see Fig.~\ref{fgcross}(c1)], and it can be located in any of the two mini-arrays, or simultaneously in both. However, we realize that there is no only one FB candidate solution for this new composed system. Interestingly, the inclusion of a connector site and the conservation of the zero amplitude site at the center of the original trimer, allow the generation of new linear localized states. In Fig.~\ref{fgcross}(c2) we show a different mode, which still has a zero amplitude at the center of the three-sites mini-array. This mode is exactly equal to the Stub FB mode~\cite{Amo}, although our composed system has additional sites below the central row. Due to symmetry, this mode can also be located in the lower row as shown in Fig.~\ref{fgcross}(c3). As a consequence, another FB mode could be the one having equal amplitudes in the top and bottom rows while having a double negative amplitude at the central connector site [see Fig.~\ref{fgcross}(c4)]. This new mode preserves the zero amplitude at the mini-array center, which indeed would be the connection of the composed system to their surrounding.

In this particular case, the inclusion of an out of axis connector site adds more complexity to the original trimer mini-array. Therefore, in order to describe all the possible FB mode candidates, a new mini-array definition is required; i.e., a four sites mini-array as the region enclosed in Fig.~\ref{fgcross}(d). Additionally to preserving the original trimer mode, now written with four components as $\{+,0,-,0 \}$, an extra mode appears with the same frequency $\beta=0$. This mode has a profile $\{+,0,0,-V_1/V_2\}$ or $\{0,0,+,-V_1/V_2\}$, where always the center site of the original three-sites mini-array remains zero. In fact, this new mode does not change our criterion for constructing FB systems. It tells us that the original three-sites mini-array was not enough to describe the new quasi-1D composed system, so we may consider a larger mini-array structure (although it allowed the generation of a 3D FB spine-like ribbon lattice as commented above).

Now, by connecting several three-sites mini-arrays using several connector sites, we are able to construct a full Cross lattice, as shown in Fig.~\ref{fgcross}(d). We define the unitary cell of this lattice [enclosed area in Fig.~\ref{fgcross}(d)], and solve the eigenvalue problem (\ref{sta}), finding that
$$\beta(k_x)=0,0,\ \pm \sqrt{4 V_2^2 \cos^2 (k_x)+2V_1^2}\ .$$
We plot these four bands in Fig.~\ref{fgcross}(e). We find two dispersive and two degenerated Flat ($\beta=0$) bands. These zero dispersion bands are composed of the localized states shown in Fig.~\ref{fgcross}(c), exactly the same obtained for the mini-array and for the composed small system [Fig.~\ref{fgcross}(b)].

\subsection{Sawtooth lattice}

\begin{figure}[t!] 
 \centering
   \includegraphics[width=.45\textwidth]{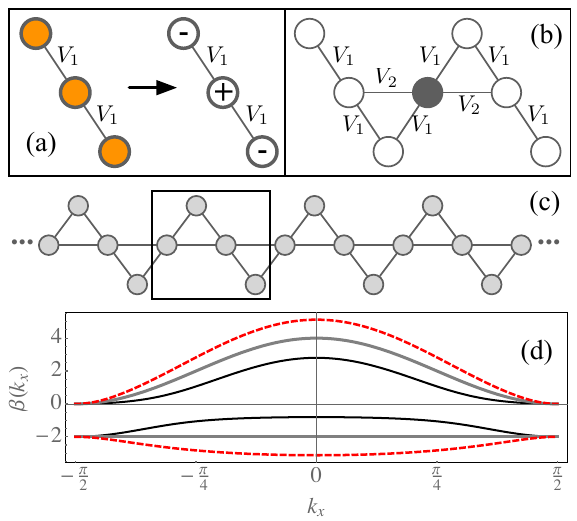}
\caption{(a) A three sites mini-array and one of its linear modes. (b) Two mini-arrays connected by a connector (dark) site. (c) Sawtooth lattice. (d) Linear spectrum for $\delta=0.75$ (thin), $\sqrt{2}$ (thick), and $2$ (dashed).}\label{fgsaw}
\end{figure}

We continue using the modes of the trimer mini-array, but this time we focus on the third one: $\{-,\sqrt{2},-\}$ [see Fig.~\ref{fgsaw}(a)]. We add a connector site and couple a second three-sites mini-array, as shown in Fig.~\ref{fgsaw}(b). As the coupling coefficient depends on the geometry, we allow the new system to have diagonal ($V_1$) and horizontal ($V_2$), in principle different, coupling coefficients. By injecting the $\{-,\sqrt{2},-\}$ mode in the first mini-array, we realize that an extra condition may be satisfied. If we write the stationary equations (\ref{sta}) for the connector site amplitude, we get
$$\beta\cdot\psi_{C} =V_1\cdot (0-A) +V_2 (\sqrt{2}A+0)=(V_2 \sqrt{2}-V_1)A\ .$$
Therefore, in order to have a zero amplitude at the connector site, a ratio $\delta\equiv V_1/V_2=\sqrt{2}$ is required. For this particular condition, the third mode of the three-sites mini-array becomes a mode of the composed system as well. In fact, it can be located in three different positions in this new composed system, because the connector site forms also a similar mini-array with a different inclination. 
%
%

By continue adding mini-arrays via connector sites, we are able to construct a full Sawtooth lattice as the one sketched in Fig.~\ref{fgsaw}(c), where the unitary cell of four sites is denoted by an enclosed area. However, due to the Sawtooth symmetry the four sites can be reduced to just two sites~\cite{SawOL}. We solve the eigenvalue problem (\ref{sta}) with this geometry and find two linear bands:
$$\beta(k_x)=V_2[\cos(2k_x) \pm f(k_x,\delta)]\ ,$$
where $f(k_x,\delta)=\sqrt{1+4 (\delta^2-1) \cos^2 (k_x)+4\cos^4 (k_x)}$. If $\delta=\sqrt{2}$, the two bands reduce to $\beta(k_x)=-2V_2,\ 4V_2\cos^2(k_x)$; i.e., a Flat Band emerges for this particular ratio between coupling coefficients. In fact, this is exactly the same condition for the third mode to be a mode of the composed system. Therefore, the $\{-,\sqrt{2},-\}$ mode is also a mode of a full Sawtooth lattice, when satisfying the condition: $\delta=\sqrt{2}$. In Fig.~\ref{fgsaw}(d) we show the linear spectrum for three different values of $\delta$, where we observe that the lower band becomes completely flat at this particular condition.

\section{2D Lieb-like examples}

\begin{figure}[h!] 
 \centering
   \includegraphics[width=.45\textwidth]{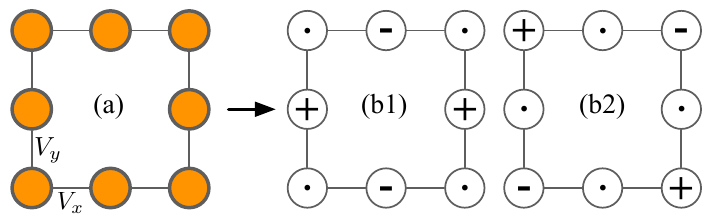}
\caption{(a) An eight sites mini-array. (b) Two linear modes.}\label{fglieb1}
\end{figure}

Now, we review our method for two-dimensional Lieb FB lattices, which have received great attention recently~\cite{njpLieb,prlus,ThomsonL,Chen3,anton2}. We start by identifying the mini-array necessary to construct a standard Lieb lattice. In Fig.~\ref{fglieb1}(a) we show a mini-array example consisting of eight sites forming a ring, including an anisotropy ($V_x\neq V_y$) degree of freedom. This system possesses eight linear modes, but we will consider only two for our study: $\{0,+,0,-V_y/V_x,-V_y/V_x,0,+,0\}$ [Fig.~\ref{fglieb1}(b1)] and $\{-,0,+,0,0,+,0,-\}$ [Fig.~\ref{fglieb1}(b2)]. These two modes are degenerated with a frequency $\beta=0$, due to a perfect cancellation of phases.

\subsection{Lieb lattice}

\begin{figure}[h!] 
 \centering
   \includegraphics[width=.4\textwidth]{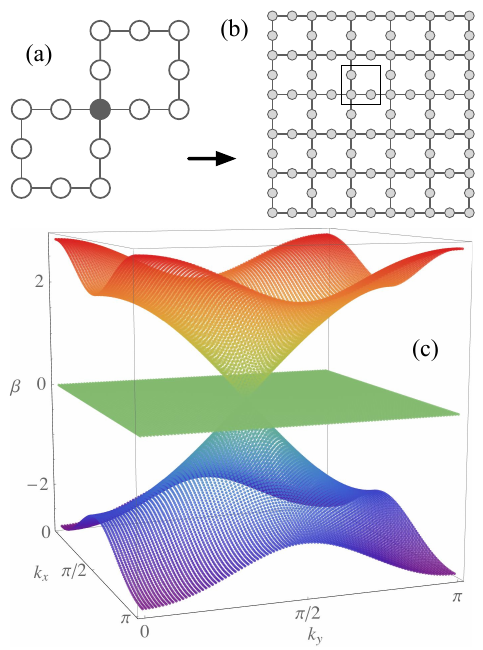}
\caption{(a) Two mini-arrays connected by a connector (dark) site. (b) Lieb lattice. (c) Linear spectrum for $V_x=V_y=1$.}\label{fglieb2}
\end{figure}

We will construct a Lieb 2D lattice by using the mini-array and first mode shown in Figs.~\ref{fglieb1}(a) and \ref{fglieb1}(b1), respectively. In this case, a zero site of the mini-array is used as a connector site (there is no need to add an extra site, although when doing it a different FB lattice could be created as we will show below). Fig.~\ref{fglieb2}(a) shows a composed system formed using the connection of two mini-arrays sharing a same connector site at one corner. As expected, the $\{0,+,0,-V_y/V_x,-V_y/V_x,0,+,0\}$ mode is also a solution of this composed system in any of the two mini-arrays, due to the perfect phase cancellation at the corner sites. By repeating this procedure in different directions, and using different corner sites, we are able to construct a full 2D Lieb lattice as shown in  Fig.~\ref{fglieb2}(b). The original mini-array mode can be located in different mini-arrays of the full lattice, all of them degenerated with $\beta=0$.
%

Solving Eqs.~(\ref{sta}) for this lattice geometry, which consider three sites per unit cell [see enclosed region in Fig.~\ref{fglieb2}(b)], we find the corresponding linear spectrum
 $$\beta(k_x,k_y)=0,\ \pm 2\sqrt{V_x^2 \cos^2 (k_x)+V_y^2\cos^2 (k_y)}\ .$$
As it is expected from our method, the degenerated mini-array modes form a complete zero-dispersion band at $\beta = 0$ [see Fig.~\ref{fglieb2}(c)]. Additionally, two dispersive bands are found for this system as well~\cite{prlus,ThomsonL}.

\subsection{Lieb 2 lattice}

\begin{figure}[h!] 
 \centering
   \includegraphics[width=.4\textwidth]{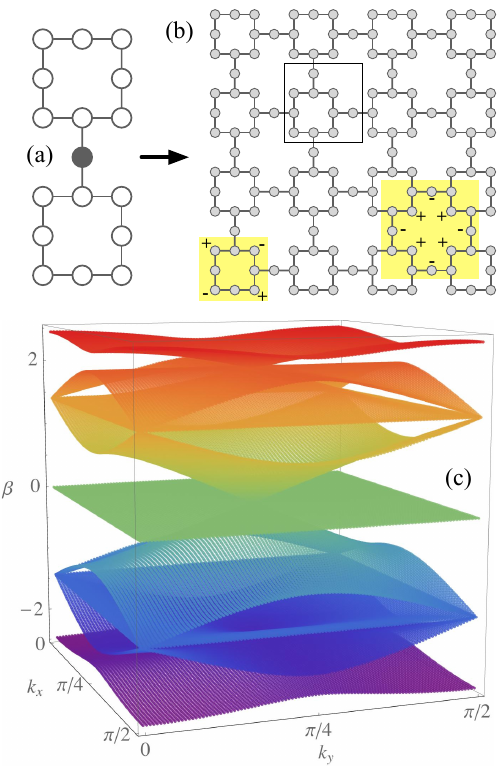}
\caption{(a) Two mini-arrays connected by a connector (dark) site. (b) Lieb 2 lattice. (c) Linear spectrum for $V_x=V_y=1$.}\label{fglieb3}
\end{figure}

We construct a ``Lieb 2'' lattice by using the mini-array and second mode shown in Figs.~\ref{fglieb1}(a) and \ref{fglieb1}(b2), respectively. For this problem, we have zero amplitude sites at the sides of the mini-array. So, we add a new connector site to connect this mini-array to another one, as shown in the example presented in Fig.~\ref{fglieb3}(a). If we inject the original stationary $\{-,0,+,0,0,+,0,-\}$ mode [Fig.~\ref{fglieb1}(b2)] in one or both mini-arrays, it will also be a solution o the composed system, because the connector site remains zero. Naturally, we can add several connector sites in the four sides of the original mini-array and connect more mini-arrays in different directions. In this way, we are able to compose a full new system that we named ``Lieb 2'' lattice, with a geometry shown in Fig.~\ref{fglieb3}(b).
%
%
This composed system possesses ten sites per unit cell [see enclosed region in Fig.~\ref{fglieb3}(b)] and, therefore, a linear spectrum having ten linear bands. The analytical form of these bands is not trivial, so we only present their plot in Fig.~\ref{fglieb3}(c). By inspecting the spectrum, we find two flat bands, both located at $\beta =0$. The first one is composed of $\{-,0,+,0,0,+,0,-\}$ mini-array modes, that continue being solutions of the extended system [see shaded area at the left-down corner of Fig.~\ref{fglieb3}(b)]. Additionally, a new localized FB state appears in the region in between four mini-arrays. It has eight sites different to zero in a staggered sign sequence, what is necessary to cancel the transport to the rest of the lattice  [see right-down shaded region in Fig.~\ref{fglieb3}(b)]. For an isotropic configuration ($V_x=V_y$), all amplitudes have equal magnitude. However, for anisotropic lattices ($V_x\neq V_y$) the relation of amplitudes becomes a bit more complicated. If we define the horizontal connector site amplitude as $C$, the vertical connector site amplitude as $B$, and the corner site amplitude as $A$, then $C=-V_yA/V_x$ and $B=-V_xA/V_y$. Similar to the Cross lattice case (Fig.~\ref{fgcross}), the appearance of this new localized state does not change our method, it only implies that the initial mini-array was not enough to describe the full new composed system.

\section{Multiple Flat Bands}

The number of FBs of a given lattice depends on the number of stationary modes that satisfy the symmetry condition at the connector sites of a given mini-array. Therefore, by analizing the specific geometry, we can construct different lattices having more than only one FB. For example, the Cross and the Lieb 2 lattices present two FBs at frequency $\beta=0$. As we discussed before, the modes belonging to these two bands are spatially different, but both preserve the condition of having a zero amplitude at the connector sites, as the original mini-array mode does. Therefore, depending on the geometry and complexity of a given mini-array, it is possible to find additional compact linear modes which are also solutions of the small system, and that preserve the condition of cancelling the transport at the same connector site.

\begin{figure}[h!] 
 \centering
   \includegraphics[width=.45\textwidth]{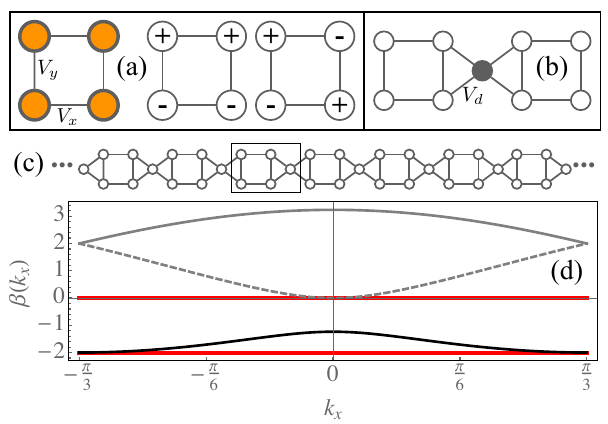}
\caption{(a) A four sites mini-array and two of their linear modes. (b) Two mini-arrays connected by a connector (dark) site. (c) B2-ribbon lattice. (d) Linear spectrum for $V_x=V_y=V_d=1$.}\label{fgrib1}
\end{figure}
%
Now, we show explicitly how to construct a system possessing more than one Flat Band. We focus on a simple system, the so-called ``B2-ribbon''. In Fig.~\ref{fgrib1}(a)-left we show a four-sites mini-array, that possesses four linear modes, with three of them being useful for FB lattice composition. If we think on an horizontally oriented ribbon lattice, the two good modes (useful for phase cancellation) are the ones shown in Fig.~\ref{fgrib1}(a)-right. These modes are denoted as $\{-,+,-,+\}$ and $\{-,+,+,-\}$, with frequencies $\beta=(V_x-V_y)$ and $\beta=-(V_x+V_y)$, respectively. We immediately observe that when connecting two mini-arrays via a connector site [as shown in Fig.~\ref{fgrib1}(b)], the dynamical equation for this site will be just zero, when initializing the system with one of these stationary modes (the connector site is symmetrically coupled to the mini-arrays with a coefficient $V_d$). By increasing the system size with more connectors and more mini-arrays, we are able to construct a full B2-ribbon lattice [see Fig.~\ref{fgrib1}(c)]. We compute the linear spectrum of this extended system, by identifying the unitary cell of this lattice [enclosed region in Fig.~\ref{fgrib1}(c)]. This cell contains five sites, therefore five linear bands are generated as Fig.~\ref{fgrib1}(d) shows. The analytical expressions for the three dispersive bands are not compact and we will not write them explicitly. The two zero dispersion bands are simply located at frequencies $\beta=(V_x-V_y)$ and $\beta=-(V_x+V_y)$, exactly at the same frequencies than the original mini-array modes. Therefore, as expected, these two modes are also localized solutions for the extended system and generate two full Flat Bands.

This example is one of the many possible configurations useful to create lattices presenting more than one FB. Of course, this method could also be extended to full 2D lattices and not only to quasi-1D ribbons. In fact, the lattice Lieb 2 (see Fig.~\ref{fglieb3}) is an example of this. In general, the dimension is not important, the key point is to preserve the discreteness of the system that allows the cancellation of phases for the modes of the mini-array and, therefore, the cancellation of transport across the lattice.

\section{Additional considerations}

\subsection{Aperiodical composition}

\begin{figure}[h!] 
 \centering
   \includegraphics[width=.47\textwidth]{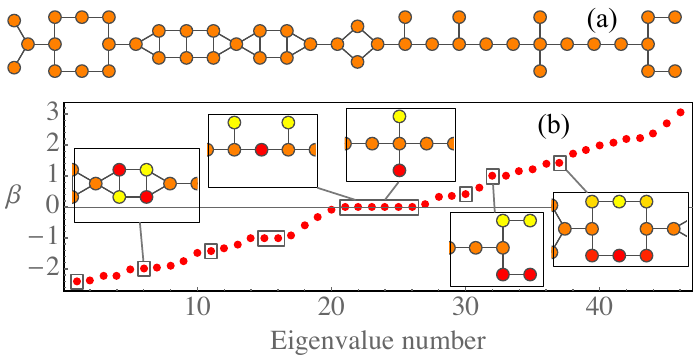}
\caption{(a) Composed aperiodic system and its (b) linear spectrum. Insets in (b) show the amplitude profile for some localized modes of this system (the color scale increases from red to yellow, with orange being equal to zero).}
\label{diso}
\end{figure}

It is possible to construct aperiodical lattices by connecting different mini-arrays via connector sites. The particular mode of every mini-array will also be a mode of the full lattice and could form a dense band, of course depending on its particular frequency. This point is very interesting because aperiodical systems will present no dispersive bands at all, but could present full FBs composed of different states. Therefore, the disorder could promote localization in the form of destructive interference of plane waves (Anderson localization~\cite{segev,morandotti1}) or due to a local geometric phase cancellation (FB localization~\cite{prlus,ThomsonL,ThomsonD}). As our criterion does not depend on the periodicity, but on the discreteness of the system (mini-array geometry), we can compose an aperiodical system as a sequence of different mini-arrays connected by several connector sites. This system will not be periodic, but still will preserve all the properties of the different coupled mini-arrays. In Fig.~\ref{diso}(a) we show an example of a small composed system, which includes eight different mini-arrays. We obtain its linear spectrum by numerically diagonalizing the corresponding coupling matrix $V_{nm}$, and plot $\beta$ versus eigenvalue number in Fig.~\ref{diso}(b). As this system is not periodic, its spectrum does not form a soft curve~\cite{disopre}. Additionally, we realize that all the mini-array modes [fourteen in this case, see boxes in Fig.~\ref{diso}(b)], that satisfy the condition for having a zero amplitude at a given connector site, are also a solution of this extended system. This array is clearly aperiodic, but can be extended to form a periodical lattice by repeating the same pattern several times. By doing this, we would generate a new periodical system, but having a more complex unitary cell [Fig.\ref{diso}(a)]. Therefore, the original mini-array states will also be a solution of the extended system and will form fourteen different degenerated Flat bands.
%

\subsection{Next-nearest neighbor coupling}

Not all the lattices preserve their zero dispersion band when including next-nearest neighbor interactions. For example, a Lieb lattice adds curvature to the original Flat Band when including diagonal coupling~\cite{anton1}, and the FB localized modes are simply lost. However, this is not an intrinsic problem of FB systems, the point is that the lattice geometry is just not the right one. In order to construct a lattice with a FB, which is robust against next-nearest neighbor interactions, there is no need to add any new ingredient to our method. We just may chose the right geometrical configuration for connector sites to assure that when considering, for example, a diagonal coupling, the dynamical equation for the connector site continues being zero, when injecting a FB mode as an initial condition. For example, this occurs for the previous Lieb 2 lattice [Fig.~\ref{fglieb3}(b)]. As we explained before, this lattice possesses two FB at $\beta=0$. When including diagonal coupling coefficients, both FB survives but one of them shifts to a frequency $\beta=-2V_d$, where $V_d$ corresponds to the next-nearest coefficient (we can see that when $V_d=0$, the band converges to its previous location). Another example is the B2-ribbon lattice (Fig.~\ref{fgrib1}). When including a diagonal interaction between vertices of the mini-array, the $\{-,+,-,+\}$ and $\{-,+,+,-\}$ modes shift their frequencies to $\beta=(V_x-V_y)-V_d$ and $\beta=-(V_x+V_y)+V_d$, respectively. Therefore, the FBs also shift their frequencies to these values and the mode profiles preserves their perfect localization. 

\subsection{Anisotropy}

In typical photonics setups, the anisotropy of crystals or of the waveguide modes is an important parameter to be taken into account when studying the linear properties of a given lattice. For example, in femtosecond written waveguide arrays~\cite{njpLieb,prlus}, the coupling interaction strongly depends on the elliptical profile of written waveguides (although the Silica buffer is essentially isotropic). On the other hand, photonic lattices induced in SBN photorefractive crystals~\cite{chen2,Chen3} also presents a strong anisotropy, but caused by the crystal itself (induced waveguides have a symmetrical profile). This consideration effectively implies that, for a fixed distance, the coupling also changes depending on their orientation ($V_x\neq V_y$)~\cite{alex2}. When computing the linear spectrum of a given lattice, the anisotropy could destroy the flatness of a previous Flat Band. For example, a Kagome lattice has no any FB when horizontal coupling differ from the diagonal one~\cite{chen2}. However, there are several systems where this effect does not affect at all the flatness of the band, for example the Cross or the Lieb lattices.

We already considered the anisotropy in our method from the very beginning, and we were able to construct robust systems presenting FBs. The only necessary condition is to have a balance between the mode amplitudes in terms of the different coupling interactions of the system. This essentially implies that the amplitude of FB modes will not have the same value at different positions. The most important consideration is that the connector dynamical equation must be zero from the beginning, when injecting a given mini-array mode. An interesting example is the Lieb 2 lattice. This system has two FBs at $\beta=0$. The first one is composed of localized modes having only four sites with amplitude different to zero [left-down corner in Fig.~\ref{fglieb3}(c)]. As we can see, the cancellation of phases for this mode is always horizontal or vertical, therefore the anisotropy balance is not required and modes simply have equal amplitudes but different phases. On the other hand, the second flat band is composed of modes which cancel the transport by balancing vertical and horizontal interactions [right-down corner in Fig.~\ref{fglieb3}(c)]. Therefore, a correction of amplitudes is required depending on a given anisotropy $V_x\neq V_y$, as we explicitly described before.

If we consider that different sites of the unitary cell could present different propagation constants (determined by $\epsilon_{\vec{n}}$), an effective anisotropy is generated in the system. Again, this is not a problem for the generation of FB lattices, however extra conditions for the balance of equations need to be fulfilled. Essentially, by starting from a given mini-array with a defined configuration of $\epsilon_{\vec{n}}$, new modes need to be computed. Then, by inspecting their symmetry one can realize if these modes are good candidates for spatial localization on an extended lattice. If the required balance is not achieved when the propagation constants are different, the coupling anisotropy could help to solve this. However, this overall balance could be too complex to be implemented in real experiments.

\subsection{Nonlinear solutions} 

One of the most common nonlinear interactions studied in diverse lattice systems corresponds to a cubic nonlinearity~\cite{rep1,rep2}. In Optics this interaction arises from the Kerr-effect, which is nothing else that an increment of the refractive index due to the intensity of a given beam. In BEC's, this interaction originates from the scattering between particles and in solid-state physics from the interaction, for example, between phonons and electrons on a given lattice. In model (\ref{DLS}), this nonlinear effect is written as $\gamma |\psi_{\vec n}|^2 \psi_{\vec n}$, where $\gamma$ corresponds to the strength of the nonlinear response. When looking for real solutions, the stationary problem to solve becomes simply
\begin{equation*}
\beta(\vec{k}) A_{\vec n} =  \sum_{\vec m \not= \vec n}V_{\vec n,\vec m} A_{\vec m}\ e^{i \vec{k} \cdot (\vec{m}-\vec{n}) }+\gamma A_{\vec n}^2 A_{\vec n}\ .
\end{equation*}
In general, any FB mode possessing a set of $N$ amplitudes ($A$), but with alternating/staggered sign, has a very simple and compact form~\cite{kag1,rib1}. As the connector sites amplitudes remain zero, the total power, defined as $P=\sum_{\vec n}|A_{\vec n}|^2$, is just given by $P=NA^2$. The frequency of the nonlinear solution becomes $\beta=\beta_0+\gamma A^2$ (where the frequency shift $\beta_0$ depends on the specific FB of a given lattice) and, therefore, a very simple and exact relation between the frequency and power arises $$P=\frac{N}{\gamma}(\beta-\beta_0)\ .$$ These nonlinear solutions are perfectly localized in a very compact spatial region. They are analytical compactons solutions~\cite{compact1,compact2}, which conserve their spatial profile in the whole range of parameters. They bifurcate at the FB position ($\beta=\beta_0$) for a zero level of power ($P = 0$). These solutions exist for positive and negative nonlinearity, with the corresponding shift on the sign frequency to assure that $P>0$.

When the FB modes possess a more complex spatial profile, including differences in the magnitude of the amplitudes, a more complicated relation is required. For example, in linear Sawtooth lattices the FB exists for a very specific value of coupling coefficients. Additionally, the FB mode has a non symmetric profile of the form $\{...,0,-1,\sqrt{2},-1,0,...\}$. Therefore, when increasing the solutions power, the amplitudes will change and the perfect balance will be more tricky. Ref.~\cite{saw1} explores how to preserve this balance in a nonlinear context (cubic and saturable), but allowing the system to also modify the coupling constants.

\subsection{Continuous model}

Finally, we study the robustness of our method in a more realistic environment. Although discrete models, based on nearest-neighbor interactions, describe very well the phenomenology observed in direct experiments~\cite{rep1,rep2}, a better prove of the stable propagation of localized FB modes is obtained by numerically solving a paraxial wave equation,
\begin{equation}
-i \frac{\partial \psi}{\partial z}= \frac{1}{2k_0n_0}\nabla_{\bot}^2 \psi + k_0\Delta n(x,y)\psi\ .
\label{model2}
\end{equation}
Here, $\psi=\psi(x,y,z)$ describes the envelope of the electric field, $k_0=2\pi/\lambda$ is the wavenumber in free space, $\lambda$ is the vacuum wavelength, and $n_0$ is the refractive index of the bulk material. The function $\Delta n(x,y)$ defines the refractive index structure, which depends on the specific lattice geometry. $\nabla_{\bot}^2 =\partial_x^2+\partial_y^2$ corresponds to the transverse Laplacian operator.

\begin{figure}[h] 
 \centering
   \includegraphics[width=.45\textwidth]{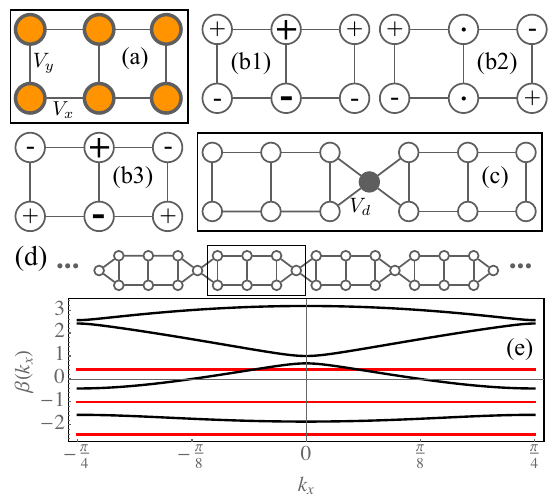}
\caption{(a) A six sites mini-array and (b) three of their linear modes. (c) Two mini-arrays connected by a connector (dark) site. (d) B3-ribbon lattice. (e) Linear spectrum for $V_x=V_y=V_d=1$.}\label{fgrib2}
\end{figure}

As an interesting example, we study a system possessing three FBs, as shown in Fig.~\ref{fgrib2}. A six-sites mini-array [Fig.~\ref{fgrib2}(a)] has three FB modes [Fig.~\ref{fgrib2}(b)] that cancel the amplitude at the connector site [Fig.~\ref{fgrib2}(c)]. The frequencies of these modes are $-V_y+\sqrt{2}V_x$, $-V_y$, and $-V_y-\sqrt{2}V_x$, respectively. When computing the linear spectrum [considering the unitary cell shown in Fig.~\ref{fgrib2}(d)], we find seven linear bands [see Fig.~\ref{fgrib2}(e)]. Four of them are dispersive and three are completely Flat. The frequencies of the zero dispersion bands are exactly the same than the ones of the mini-array modes shown in Fig.~\ref{fgrib2}(b), as expected from our method.
%
%

To study this lattice in a more realistic configuration, we insert the lattice geometry in model (\ref{model2}) by defining the function $\Delta n(x,y)$~\cite{prlus}. We assume elliptical waveguides~\cite{alex1}, what induces a strong effective anisotropy: $V_y\neq V_x$. This does not affect the flatness of the three degenerated bands, it only implies a different balance between the FB modes amplitudes. Considering standard parameters used in Ref.~\cite{prlus} ($n_0=1.40$, lattice period of $d=20\ \mu$m, propagation distance of $L=10$ cm, $\lambda= 532$ nm, and maximum index contrast of $0.67\times10^{-3}$), we implement a Beam Propagation Method to solve Eq.(\ref{model2}) numerically.

\begin{figure}[h] 
 \centering
   \includegraphics[width=.4\textwidth]{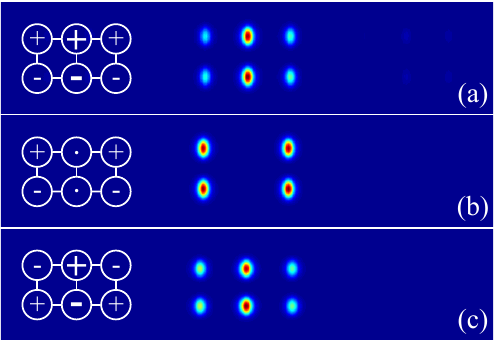}
\caption{Transversal intensity profile $|\psi(x,y,L)|^2$, after propagating a distance $L=10$ cm, in a B3-ribbon lattice geometry. (a)-(c) Correspond to different output profiles, for the input conditions sketched on each figure.}\label{fgcont}
\end{figure}

In Fig.~\ref{fgcont} we show our results for three different input conditions. We observe that the three input profiles, corresponding to the three FB modes of a B3-ribbon lattice [see Fig.~\ref{fgrib2}(b)], propagate a long distance without suffering noticeable distortion. In fact, Fig.~\ref{fgcont}(a) shows the excitation of some outsider lobes, which are very weak in comparison to the central highly excited sites. In Fig.~\ref{fgcont}(b) and (c) only the mode profiles are observable, with essentially a zero background. This result is certainly very interesting because we are using discrete solutions as input conditions of a continuous model. It is well known that a discrete (tight-binding) model describes only the lower part of the spectrum of an, in principle, infinite band-gap system~\cite{kittel}. When thinking on corrections to the nearest-neighbors model (\ref{DLS}), the first consideration to have in mind are diagonal or next-nearest neighbor interactions. The B3-ribbon lattice example is robust against to anisotropy and, also, to second order linear interactions. This is a good indication that the discrete predictions will be observable for long distances in a continuous (realistic) medium. However, as this is an approximation of a more complex system, which includes an infinite set of linear bands, after even longer distances the FB states may start to experience diffraction across the lattice. 
%
%

\section{Conclusions}

In this work, we developed a simple method to construct different FB lattices. By inspecting the mini-array mode profiles, we identify the good candidate modes for cancelling the transport across a full lattice, as a consequence of a local geometric phase cancellation at the connector sites. Our method is based on the discrete properties of a given lattice and the right identification of the corresponding mini-array. In this way, our method is able to explain all the already known FB lattice systems, but also give a simple receipt for inventing new ones. As our technique is based on the modes of a given mini-array, the lattice composition method will always give a complete FB, composed of spatially localized linear modes. By allowing the system to have extra interactions, we can also find robust lattices which preserve the localization of FB modes when including, for example, next nearest-neighbor interactions. This becomes very important when studying the lattice phenomenology experimentally, as we showed by numerically solving a continuous model. Finally, we showed that some nonlinear FB lattices possess simple analytical compact solutions, which could be relevant when thinking on the mobility of strongly localized wave-packets~\cite{kag1}.

\section*{Acknowledgements}

Authors acknowledge useful discussions with C. Cantillano and B. Real. The authors wish to thank Programa ICM P10-030-F and FONDECYT Grant No. 1151444.


\begin{thebibliography}{99}

\bibitem{kittel}C. Kittel, ``Introduction to Solid State Physics,'' John Wiley and Sons, Inc., seventh edition (1996).

\bibitem{rep1} F. Lederer, G.I. Stegeman, D.N. Christodoulides, G. Assanto, M. Segev, and Y. Silberberg, ``Discrete solitons in optics,'' Phys. Rep. \textbf{463}, 1 (2008).

\bibitem{anderson}P.W. Anderson, Phys. Rev. {\bf 109}, 1492 (1958).

\bibitem{segev}T. Schwartz, G. Bartal, S. Fishman and M. Segev, Nature \textbf{446}, 52-55 (2007).

\bibitem{morandotti1}Y. Lahini, A. Avidan, F. Pozzi, M. Sorel, R. Morandotti, D. N. Christodoulides, and Y. Silberberg, \prl \textbf{100}, 013906 (2008).

\bibitem{morandotti2}H.S. Eisenberg, Y. Silberberg, R. Morandotti, A.R. Boyd, and J.S. Aitchison, \prl \textbf{81}, 3383 (1998).

\bibitem{segev2}J.W. Fleischer, T. Carmon, M. Segev, N.K. Efremidis, and D.N. Christodoulides, \prl \textbf{90}, 023902 (2003).

\bibitem{alex1}A. Szameit, D. Bl\"omer, J. Burghoff, T. Schreiber, T. Pertsch, S. Nolte, A. T\"unnermann, and F. Lederer, Opt. Express \textbf{13}, 10552 (2005).

\bibitem{segev3}J.W. Fleischer, M. Segev, N.K. Efremidis, and D.N. Christodoulides, Nature \textbf{422}, 147 (2003).

\bibitem{alex2Ds}A. Szameit, J. Burghoff, T. Schreiber, T. Pertsch, S. Nolte, A. T\"unnermann, and F. Lederer, Opt. Exp. \textbf{14}, 6055 (2006).

\bibitem{PT04}D.K. Cambpell, S. Flach, and Y.S. Kivshar, Phys. Today \textbf{57}, 43 (2004).

\bibitem{rep2}S. Flach and A. Gorbach, Phys. Rep. {\bf 467}, 1 (2008).

\bibitem{Chen1} F. Fedele, J. Yang, and Z. Chen, Opt. Lett. {\bf 30}, 1506 (2005).

\bibitem{Chen1a} I. Makasyuk, Z. Chen, and J. Yang, Phys. Rev. Lett. {\bf 96}, 223903 (2006).

\bibitem{ThomsonD}S. Mukherjee and R.R. Thomson, Opt. Lett.  {\bf 40}, 5443 (2015).

\bibitem{Amo}F. Baboux, L. Ge, T. Jacqmin, M. Biondi, A. Lema\^{\i}tre, L. Le Gratiet, I. Sagnes, S. Schmidt, H.E. T\"ureci, A. Amo, J. Bloch, Phys. Rev. Lett.  {\bf 116}, 066402 (2016).

\bibitem{chi1}T. Zhang and G.-B. Jo, Sci. Rep. {\bf 5}, 16044 (2015).

\bibitem{SawOL}S. Weimann, L. Morales-Inostroza, B. Real, C. Cantillano, A. Szameit, and R.A. Vicencio, Opt. Lett. {\bf 41}, 2414 (2016).

\bibitem{stan1}S.A. Parameswaran, I. Kimchi, A.M. Turner, D.M. Stamper-Kurn, and A. Vishwanath, Phys. Rev. Lett. {\bf 110}, 125301 (2013).

\bibitem{chen2} Y. Zong, S. Xia, L. Tang, D. Song, Y. Hu, Y. Pei, J. Su, Y. Li, and Z. Chen, Optics Express {\bf 24}, 8877 (2016).

\bibitem{prlus}R.A. Vicencio, C. Cantillano, L. Morales-Inostroza, B. Real, C. Mej\'ia-Cort\'es, S. Weimann, A. Szameit, and M.I. Molina, Phys. Rev. Lett. {\bf 114}, 245503 (2015).

\bibitem{ThomsonL}S. Mukherjee, A. Spracklen, D. Choudhury, N. Goldman, P. \"Ohberg, E. Andersson, and R.R. Thomson, Phys. Rev. Lett. {\bf 114}, 245504 (2015).

\bibitem{sci1}S. Taie, H. Ozawa, T Ichinose, T. Nishio, S. Nakajima, Shuta and Y. Takahashi, Sci. Adv. {\bf 1}, e1500854 (2015). 

\bibitem{kagome1}R.A. Vicencio and C. Mej\'ia-Cort\'es, J. Opt. {\bf 16}, 015706 (2014).

\bibitem{Chen3}S. Xia, Y. Hu, D. Song, Y. Zong, L. Tang, and Z. Chen, Opt. Lett. 41, {\bf 1435} (2016).

\bibitem{origin}S. Deng, A. Simon, and J. K\"ohler, J. of Solid State Chemistry {\bf 176}, 412 (2003).

\bibitem{japan}S. Miyahara, K. Kubo, H. Ono, Y. Shimomura, N. Furukawa, Journal of the Physical Society of Japan {\bf 74}, 1918 (2005).

\bibitem{bergman}D.L. Bergman, C. Wu, and L. Balents, Phys. Rev. B {\bf 78}, 125104 (2008).

\bibitem{flach}S. Flach, D. Leykam, J.D. Bodyfelt, P. Matthies, and A.S. Desyatnikov, Eur. Phys. Lett. {\bf 105}, 30001 (2014).

\bibitem{manni}M. Hyrk\"as, V. Apaja, and M. Manninen, Phys. Rev. A {\bf 87}, 023614 (2013).

\bibitem{alex2}A. Szameit, F. Dreisow, T. Pertsch, S. Nolte, and A. T\"unnermann, Opt. Exp. {\bf 15}, 1579 (2007).

\bibitem{anton2}F. Diebel, D. Leykam, S. Kroesen, C. Denz, and A.S. Desyatnikov, \prl\ {\bf 116}, 183902 (2016).

\bibitem{unisign}R. Keil, C. Poli, M. Heinrich, J. Arkinstall, G. Weihs, H. Schomerus, and A. Szameit, Phys. Rev. Lett. \textbf{116}, 213901 (2016).

\bibitem{njpLieb}D Guzm\'an-Silva, C. Mej\'ia-Cort\'es, M.A. Bandres, M.C. Rechtsman, S. Weimann, S. Nolte, M. Segev, A. Szameit and R.A. Vicencio, New J. Phys. \text{16}, 063061 (2014).

\bibitem{disopre}R.A. Vicencio and S. Flach, \pre \textbf{79}, 016217 (2009).

\bibitem{anton1}D. Leykam, O. Bahat-Treidel, A.S. Desyatnikov, \pra\ {\bf 86}, 031805(R) (2012).

\bibitem{kag1}R.A. Vicencio and M. Johansson, Phys. Rev. A \textbf{87} 061803(R) (2013).

\bibitem{rib1}P.P. Beli\v{c}ev, G. Glirori\'c, A. Radosavljevi\'c, A. Maluckov, M. Stepi\'c, R.A. Vicencio, and M. Johansson, Phys. Rev. E \textbf{92}, 052916 (2015).

\bibitem{compact1}P.G. Kevrekidis and V.V. Konotop, Phys. Rev. E \textbf{65}, 066614 (2002).

\bibitem{compact2}P.G. Kevrekidis, V.V. Konotop, A.R. Bishop, and S. Takeno, J. Phys. A \textbf{35}, L641 (2002).

\bibitem{saw1}M. Johansson, U. Naether, and R.A. Vicencio, Phys. Rev. E \textbf{92}, 032912 (2015).

\end{thebibliography}
\end{document}